# Control of graphene's properties by reversible hydrogenation

D. C. Elias[1], R. R. Nair[1], T. M. G. Mohiuddin[1], S. V. Morozov[2], P. Blake[3], M. P. Halsall[1], A. C. Ferrari[4], D. W. Boukhvalov[5], M. I. Katsnelson[5], A. K. Geim[1,3], K. S. Novoselov[1]

[1]School of Physics & Astronomy, University of Manchester, M13 9PL, Manchester, UK
[2]Institute for Microelectronics Technology, 142432 Chernogolovka, Russia
[3]Manchester Centre for Mesoscience & Nanotechnology, University of Manchester, M13 9PL, Manchester, UK
[4]Department of Engineering, Cambridge University, 9 JJ Thomson Avenue, Cambridge CB3 OFA, UK
[5]Institute for Molecules and Materials, Radboud University Nijmegen, 6525 ED Nijmegen, The Netherlands

**Graphene – a monolayer of carbon atoms densely packed into a hexagonal lattice *(1)* –has one of the strongest possible atomic bonds and can be viewed as a robust atomic-scale scaffold, to which other chemical species can be attached without destroying it. This notion of graphene as a giant flat molecule that can be altered chemically is supported by the observation of so-called graphene oxide that is graphene densely covered with hydroxyl and other groups *(2,3,4,5,6)*. Unfortunately, graphene oxide is strongly disordered, poorly conductive and difficult to reduce to the original state *(6)*. Nevertheless, one can imagine atoms or molecules being attached to the atomic scaffold in a strictly periodic manner, which should result in a different electronic structure and, essentially, a different crystalline material. A hypothetical example for this is graphane *(7)*, a wide-gap semiconductor, in which hydrogen is bonded to each carbon site of graphene. Here we show that by exposing graphene to atomic hydrogen, it is possible to transform this highly-conductive semimetal into an insulator. Transmission electron microscopy reveals that the material retains the hexagonal lattice but its period becomes markedly shorter than that of graphene, providing direct evidence for a new graphene-based derivative. The reaction with hydrogen is found to be reversible so that the original metallic state and lattice spacing are restored by annealing and even the quantum Hall effect recovers. Our work proves the concept of chemical modification of graphene, which promises a whole range of new two-dimensional crystals with designed electronic and other properties.**

Graphene continues to attract immense interest, mostly due to its unusual electronic properties, strictly two-dimensional nature and atomic thickness *(1)*. The venue that so far remains practically unexplored is chemical modification of graphene. Indeed, the ability of carbon to change its coordination number and chemical bonding, which is known to result in several allotropes and myriads of carbon-based compounds, can arguably be exploited to create chemical derivatives of graphene as well. This would widen the range of electronic, chemical, mechanical and other properties achievable in graphene-based materials and make them better suitable for specific applications.

Particularly simple and elegant is the idea of attaching atomic hydrogen to each site of the graphene lattice, which changes the hybridisation from $sp^2$ into $sp^3$, thus removing the conducting π-bands and opening an energy gap *(7,8)*. Previously, absorption of hydrogen on graphitic surfaces was investigated mostly in conjunction with hydrogen storage, with the research focused on physisorbed molecular hydrogen *(9,10,11)*. More recently, atomic hydrogen chemisorbed on carbon nanotubes has been studied theoretically *(12)* as well as by a variety of experimental techniques including infrared *(13)*, ultraviolet *(14,15)*, x-ray *(16)* and STM *(17)* spectroscopy. In this work, we report on reversible hydrogenation of single-layer graphene, which results in dramatic changes in its properties as observed by transport measurements, Raman spectroscopy and transmission electron microscopy (TEM).

Graphene crystals studied in this work were prepared by micromechanical cleavage *(18)* of graphite on top of an oxidized Si substrate (300nm $SiO_2$) and then identified by their optical contrast *(1,18)* and distinctive Raman signatures *(19)*. Three types of samples were used: large (>20 μm) crystals for Raman studies, the standard Hall bar devices of 1 μm in width *(18)*, and free-standing membranes *(20,21)* for TEM. For details of sample fabrication we refer to earlier work *(18,20,21)*. All samples were first annealed at 300°C in argon atmosphere for 4 hours in order to remove any possible contamination (e.g, adsorbed water and resist residues). After their initial characterization, the samples were exposed to cold hydrogen plasma. We used a low pressure (1 mbar) hydrogen-argon mixture (10% $H_2$) with *dc* plasma ignited between two aluminium electrodes. The samples were kept 30 cm away from the discharge zone, in order to minimize any possible damage by energetic ions. We found that it typically required 2 hours of plasma treatment for reaching the saturation in measured characteristics.

Figure 1 illustrates typical changes induced by the hydrogenation in electronic properties of graphene. Before their plasma exposure, our devices exhibited the standard ambipolar field effect with the neutrality point (NP) close to zero gate voltage *(18)*. For the device shown in Fig. 1A, mobility $\mu$ of charge carriers was $\approx$14,000 cm$^2$/V·s. The device showed a very weak temperature dependence that became metallic below 50K *(22)* and the half-integer quantum Hall effect (QHE) (Fig. 1B), which both are landmarks of single-layer graphene (see Ref. *(1)* and references therein). This behaviour completely changed after the devices were treated with atomic hydrogen (Figs. 1C,D). They started exhibiting an insulating behaviour such that resistivity $\rho$ grew by two orders of magnitude with decreasing temperature *T* from 300 to 4 K (Fig. 1C). This was accompanied by a decrease in low-*T* mobility down to values of ~10 cm$^2$/V·s for typical carrier concentrations *n* of the order of 10$^{12}$ cm$^{-2}$. The quantum Hall plateaus, so marked in the original devices, completely disappeared, with only weak signatures of Shubnikov-de-Haas oscillations remaining in magnetic field *B* of 14 T (Fig. 1D). In addition, we observed a shift of NP to gate voltages $V_g \approx$+50V, which showed that graphene became doped with holes in concentration $\approx$3.5$\times$10$^{12}$ cm$^{-2}$. Not too far from NP, the observed temperature dependences $\rho(T)$ can be well fitted by the function $\exp[(T_0/T)^{1/3}]$ (Fig. 2), which is a clear signature of variable-range hopping in two dimensions *(23)* ($T_0$ is the parameter that depends on $V_g$). $T_0$ exhibits a maximum of ~260K at NP and strongly decreases away from NP (Fig. 2B). At *n* >4$\times$10$^{12}$ cm$^{-2}$ (both for electrons and holes), changes in $\rho$ with *T* became small (similar to those in pristine graphene), which indicates a transition from the insulating to metallic regime.

The hydrogenated devices were stable at room *T* for many weeks showing the same characteristics during repeated measurements. However, we found that it was possible to restore the original metallic state by annealing (we used 450$^o$C in Ar atmosphere for 24 hours; higher annealing *T* damaged graphene). After the annealing, the devices returned practically to the same state as before hydrogenation: $\rho$ as a function of $V_g$ reached again a maximum value of $\approx h/4e^2$ and became only weakly $T$ dependent (Figs. 1E and 2). Also, $\mu$ recovered to ~3,500 cm$^2$/V·s, and the QHE reappeared (Fig. 1F). Still, the recovery was not complete: Graphene remained p-doped, the QHE did not restore at filling factors $\nu$ larger than $\pm 2$ (cf. Figs. 1B and 1F), and zero-$B$ conductivity $\sigma = 1/\rho$ became a sublinear function of $n$ (not shown). We attribute the remnant features to vacancies induced by plasma damage. To this end, note that after annealing the distance (as a function of $V_g$) between the peaks in $\rho_{xx}$ at $\nu = 0$ and $\nu = \pm 4$ became notably (~40%) larger than that between all the other peaks for both annealed and original devices. The larger distance indicates mid-gap states *(24)* (such as vacancies *(25)*) induced during the processing, in agreement with the observed sublinear behaviour. The extra charge required to fill these states yields *(24)* their density as of about $1 \times 10^{12}$ cm$^{-2}$ (average distance of $\approx 10$ nm).

The significant changes induced by hydrogenation have been corroborated by Raman spectroscopy. The main features in the Raman spectra of carbon-based materials are the G and D peaks that lie at around 1580 and 1350 cm$^{-1}$, respectively. The G peak corresponds to optical $E_{2g}$ phonons at the Brillouin zone centre, whereas the D peak is due to the breathing-like modes (corresponding to TO phonons close to the K point) and requires a defect for its activation via an inter-valley double resonance Raman process *(19,26,27,28)*. Both the G and D peaks are due to vibrations of sp$^2$ hybridised carbon atoms. The D peak intensity provides a convenient measure for the amount of disorder in graphene *(27,28)*. Its overtone, the 2D peak, appears around 2680cm$^{-1}$, fingerprinting monolayer graphene (19) , and is present even in the absence of any defects, being the sum of two phonons with opposite momentum. In Fig. 3 one can also notice a peak at ~1620 cm$^{-1}$, called D', which occurs via an intra-valley double resonance process in the presence of defects.

Fig. 3A shows the evolution of Raman spectra for graphene crystals hydrogenated and annealed simultaneously with the device in Fig. 1 (the use of different samples for Raman studies was essential to avoid an obscuring contribution due to edges of the Hall bars, that were smaller than our laser spot size of about a µm). Hydrogenation resulted in the appearance of sharp D and D' peaks, slight broadening and a decrease of the height of the 2D peak relative to the G peak, and the onset of a combination mode (D+D') around 2950 cm$^{-1}$, which, unlike the 2D and 2D' bands, requires a defect for its activation, since it is a combination of two phonons with different momentum. The D peak in hydrogenated graphene is observed at 1342 cm$^{-1}$ and is very sharp, as compared to that in disordered or nanostructured carbon-based materials *(28)*. We attribute the development of this sharp D peak in our hydrogenated samples to interruptions of the π electron delocalisation consequent to the formation of C-H sp$^3$ bonds.

After annealing, the Raman spectrum recovered to almost its original shape so that all the defect-related peaks (D, D' and D+D') became strongly suppressed. Two broad low intensity bands appear, overlapping a sharper G and residual D peaks. These are indicative of some residual structural disorder (*28*). The 2D peak remained relatively small with respect to the G peak, when compared to the same ratio in the pristine sample, and both became shifted to higher energies yielding that the annealed graphene is p-doped *(29)*. The observed changes are in broad agreement with our transport experiments.

For graphene on a substrate, only one side is accessible to atomic hydrogen, and the plasma exposure is not expected to result in graphane (which assumes hydrogen atoms attached on both sides). For more effective hydrogenation, we employed free-standing graphene samples (*20,21*) (membranes) (Fig. 3B; inset). Raman spectra for hydrogenated and consequently annealed membranes (Fig. 3B) were rather similar to those described above for graphene on SiO$_2$, but with some notable differences. If hydrogenated simultaneously and before reaching the saturation, the D peak for a membrane was by a factor of 2 larger than that for graphene on a substrate (Fig. 3A; inset), which indicates the formation of a double amount of C-H bonds in the membrane. This agrees with the general expectation that atomic hydrogen attaches to both sides of membranes. Moreover, the D peak could become notably (up to 2.5 times) larger than the G peak after prolonged exposures to atomic hydrogen (Fig. 3B).

Further information about hydrogenated membranes was obtained by TEM (we used Tecnai F30). For graphene, the electron diffraction patterns observed on dozens of the studied membranes were always the same, exhibiting the hexagonal symmetry with the lattice constant $d = 2.46 \pm 0.02$ Å. Prolonged exposure to atomic hydrogen was found to result in drastic changes in the crystal structure such that $d$ shrunk by as much as 5% whereas the hexagonal symmetry remained clearly preserved. Fig. 4A shows an example of the strongly compressed lattice. Generally, the compression was not uniform and different parts of membranes exhibited locally different in-plane periodicities (see Fig. 4B; diameters of the selected area for the electron diffraction and studied membranes were 0.3 µm and 30 to 50 µm, respectively). Such non-uniformity is not unexpected because the crystals were fixed to a metal scaffold (Fig. 3) and, therefore, could not possibly shrink isotropically. To this end, we also note that partially broken membranes (with extended free edges) such as shown in Fig. 3A were hydrogenated more uniformly than membranes covering the whole aperture (see Supplementary Information). The annealing led to complete recovery of the original periodicity.

The in-plane compression of graphene's lattice is impossible by any other means except for chemical modification, because any compression that is not stabilized on an atomic scale should lead to membrane's buckling. The most obvious candidate for the modified crystal lattice is graphane *(7,8)*. In this until-now-theoretical material, hydrogen attaches to graphene's sublattices A and B from the two opposite sides and carbon atoms in A and B move out of the plane ("buckle") as shown in Fig. 4D. The in-plane periodicity probed by TEM would then shrink significantly if the length $a$ of the C-C bond were to remain the same as in graphene (1.42 Å). However, the change in hybridization from $sp^2$ to $sp^3$ generally results in longer C-C bonds, which is the effect opposing to the lattice shrinkage by atomic-scale buckling. Recent calculations *(8)* predicted $a$ in graphane to be ≈1.53 Å (close to that in diamond) and the in-plane periodicity $d$ of ≈1% smaller than in graphene. Although the maximum in the observed distribution of $d$ in Fig. 4B occurs at ≈2.42 Å (Fig. 4B), i.e. close to the theoretical value for graphane, the observation of more compressed areas (such as in Fig. 4A) suggests that the equilibrium $d$ (that is without strain imposed by the scaffold) should be smaller. This implies either shorter or stronger buckled C-C bonds or both. Alternatively, the experimentally produced graphane may have a more complex hydrogen bonding than the one suggested by theory.

Finally, let us return to the graphene hydrogenated on a substrate (Fig. 1 and 3). It has been shown that single-sided hydrogenation is thermodynamically unstable *(7,8)* and, therefore, our experiments seem to be in conflict with the theory. However, we note that realistic graphene samples are not microscopically flat but always rippled *(20,21)*, and this should facilitate single-sided hydrogenation. Indeed, attached hydrogen is expected to change the hybridisation of carbon from $sp^2$ to (practically) $sp^3$ with angles of $\sim 110^o$ acquired between all the bonds *(7)*. This necessitates carbon atoms to move out of the plane in the direction of the attached hydrogen, which costs elastic energy. However, for a convex surface, the lattice is already deformed in the direction that favours $sp^3$ bonding, which lowers the total energy. As shown in the Supplementary Information, single-sided hydrogenation becomes energetically favourable for a typical size of ripples observed experimentally *(20)*. Because of the random nature of ripples, single-sided graphane is expected to be a disordered material, similar to graphene oxide, rather than a new graphene-based crystal. This explains with the observation of variable-range hopping in our transport experiments.

In conclusion, the distinct crystal structure of hydrogenated graphene supported by the observation of pronounced changes in its electronic and phonon properties unambiguously indicate new graphene derivatives (crystalline and disordered for the case of double- and single- sided hydrogenation, respectively). The results prove that novel graphene-based molecules with a well-defined atomic structure are indeed possible. The reversible nature of graphene's hydrogenation points at enticing prospects of using this effect in hydrogen storage.

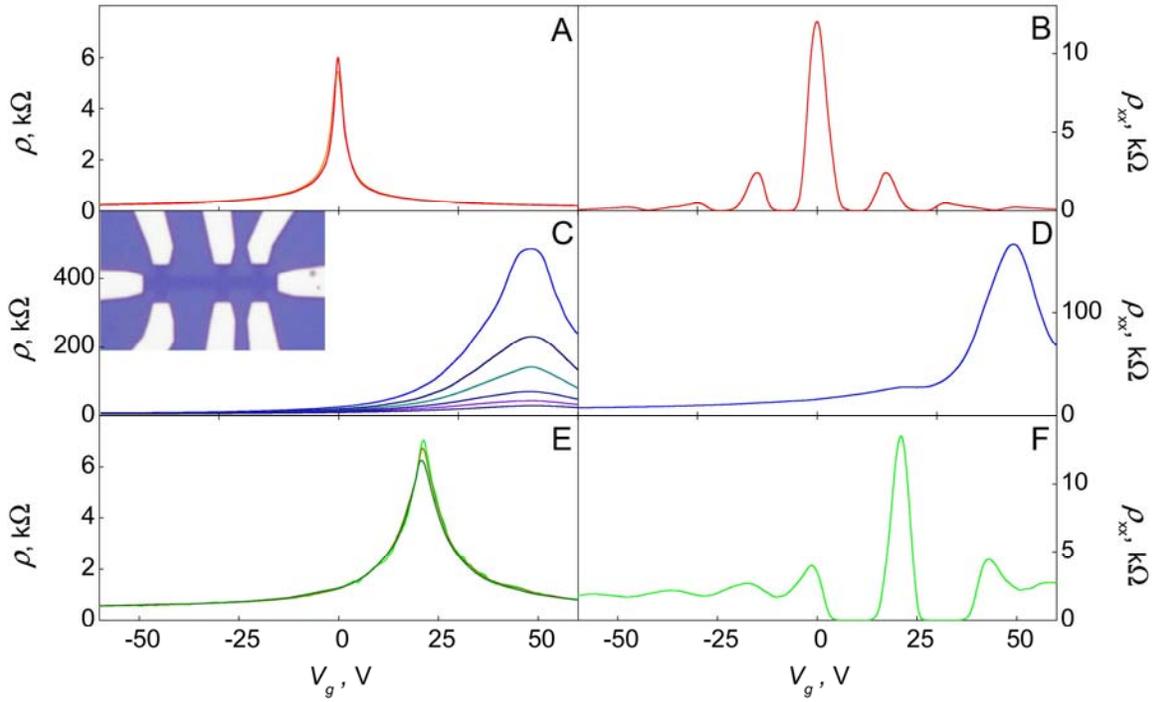

Figure 1. Control of the electronic properties of graphene by hydrogenation. The electric field effect for one of our devices in zero $B$ at various $T$ (left column) and in $B$ =14 T at 4 K (right). (A,B) - The sample before its exposure to atomic hydrogen; curves in (A) are for three different $T$ of 40, 80 and 160K and practically coincide. (C,D) – After hydrogen treatment. In (C), $T$ =4, 10, 20, 40, 80 and 160 K from top to bottom. (E,F) – The same sample after annealing. (E) - $T$ =40, 80 and 160 K. Inset: Optical micrograph of our typical Hall bar device. The scale is given by its width of 1 μm.

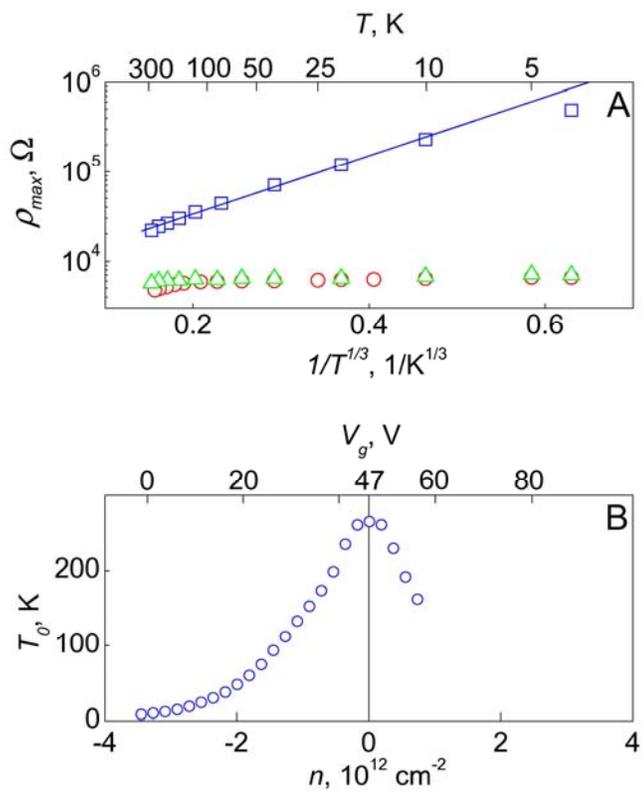

Figure 2. Metal-insulator transition in hydrogenated graphene. (A) - Temperature dependence of graphene's resistivity at NP for the sample shown in Fig. 1. Red circles, blue squares and green triangles are for pristine, hydrogenated and annealed graphene, respectively. The solid line is a fit by the variable-range hopping dependence $\exp[(T_0/T)^{1/3}]$. (B) - Characteristic exponents $T_0$ found from this fitting at different carrier concentrations.

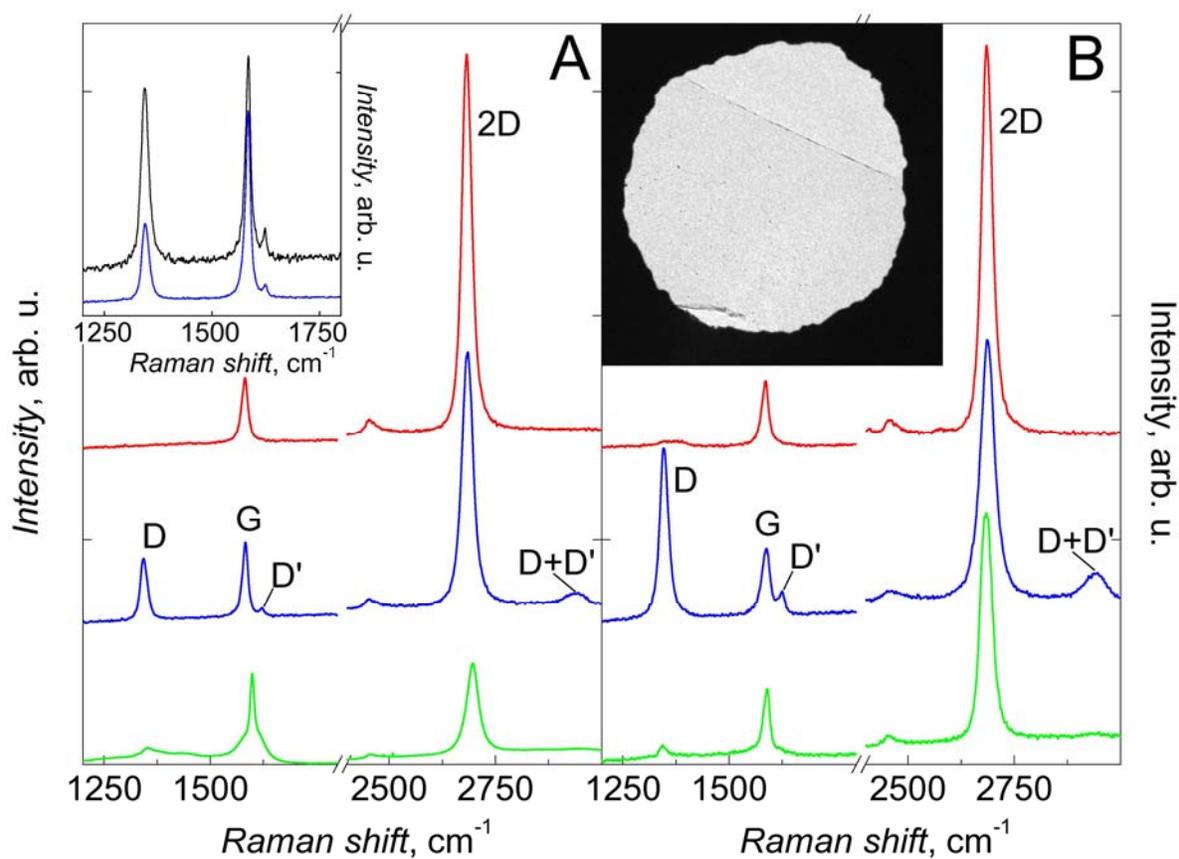

Figure 3. Changes in Raman spectra of graphene due to its hydrogenation. (A) – graphene on $SiO_2$; (B) – free-standing graphene. Red, blue and green curves correspond to pristine, hydrogenated and annealed samples, respectively. Graphene was hydrogenated for ~3 hours, and the spectra were measured with a Renishaw spectrometer at wavelength 514 nm and low power (1.2 mW; to avoid damage during measurements). Left inset: Comparison between the evolution of D and D' peaks for single- and double-sided exposure to atomic hydrogen. Shown is a partially hydrogenated state achieved after 1 hour of simultaneous exposure of graphene on $SiO_2$ (blue curve) and of a membrane (black). Right inset: TEM image of one of our membranes that partially covers the aperture of 50 μm in diameter.

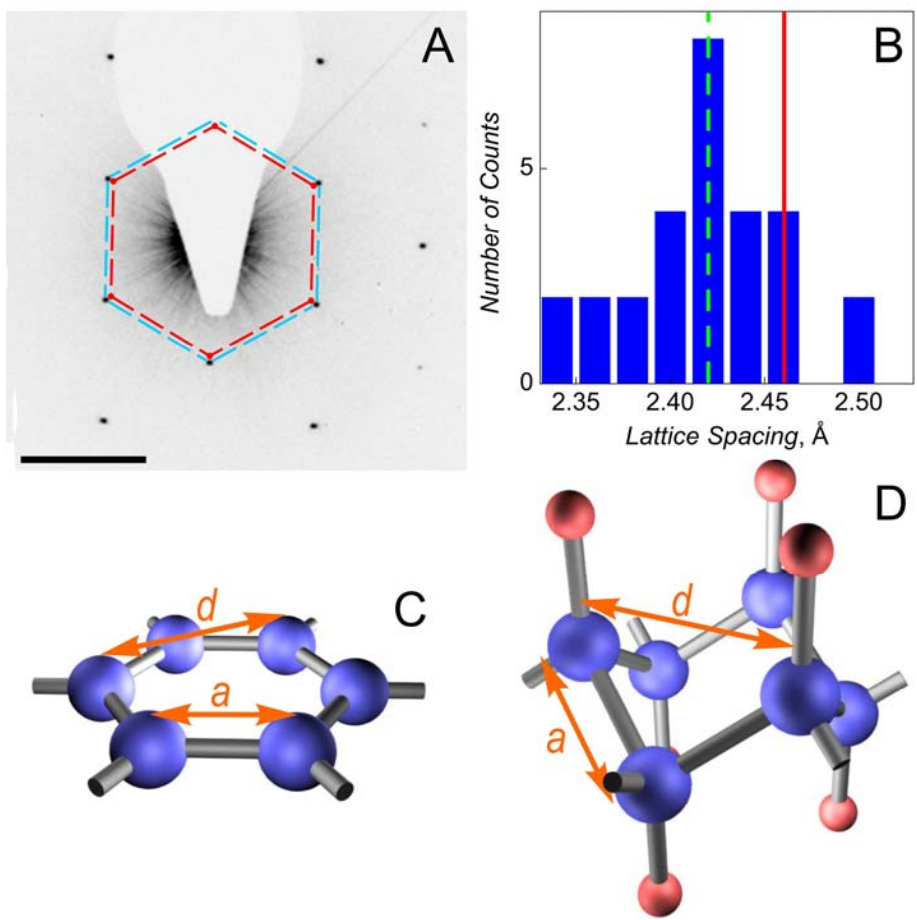

Figure 4. Graphane: crystalline derivative of graphene. (A) – Changes in the electron diffraction after ~4h exposure of graphene membranes to atomic hydrogen. The scale bar is 5 nm$^{-1}$. The blue hexagon is a guide for the eye and marks positions of the diffraction spots in graphane. The equivalent diffraction spots in graphene under the same conditions are shown by the red dots and hexagon. (B) – Distribution of the lattice spacing *d* found in hydrogenated graphene membranes. The green line marks the average value whereas the red one shows *d* always observed for graphene (i.e. both before hydrogenation and after annealing). (C,D) - Schematic representation of the crystal structure of graphene and theoretically predicted graphane. Carbon (hydrogen) atoms are shown as blue (red) spheres.

# Supporting Information

"Control of graphene's properties by reversible hydrogenation" by D. C. Elias *et al*

1. Non-uniform Hydrogenation of Graphene Membranes

In the main text, we describe graphane crystals exhibiting lattice periodicity $d$ notably shorter than that in graphene. We have observed a range of $d$ values rather than a single period expected for homogenous graphane. This was attributed to the inevitable mechanical strain induced when parts of the crystal rigidly attached to a metal scaffold had to shrink during conversion of graphene into graphane. To minimize the effect of strain, we chose to work with partially broken membranes that allowed at least a part of the strain to relax due to the presence of free boundaries.

To emphasize the importance of strain effects, here we describe what happens in the extreme case when a graphene membrane has no free boundaries. Figure S1A shows a TEM micrograph of a hydrogenated crystal that completely covers a 50 μm aperture in a 15 μm thick copper film *(S1)*. For this membrane, we observed not only the regions that contracted but also those that drastically expanded. The former exhibited $d$ down to 2.37±0.02 Å (i.e. 4% smaller than in graphene), in agreement with the results for partial and ruptured membranes reported in the main text. In the expanded regions, $d$ could be as large as ≈2.7 Å (i.e. the lattice was stretched isotropically by nearly 10% with respect to pristine graphene). Figure S1B shows an example of a stretched graphene lattice. One can see that the diffraction spots occur inside the red hexagon indicating the diffraction pattern in graphene, which is directly opposite to Fig. 4A that shows the same diffraction spots but outside the red hexagon. This amount of stretching is close to the limit of possible elastic deformations in graphene *(S2)* and, indeed, we observed some of our membranes to rupture during their hydrogenation. We believe that the stretched regions are likely to remain non-hydrogenated.

It is hardly surprising that if a part of a rigidly fixed membrane shrinks, other parts have to extend. However, we found that instead of exhibiting random stretching, graphene membranes normally split into domain-like regions. The symmetry within each domain remained hexagonal but with either increased or decreased *d*. In other words, not only contraction due to hydrogenation was isotropic (as already discussed in the main text) but also the expansion was mostly isotropic. Having said that, we also observed expanded regions, for which the hexagonal diffraction pattern was stretched along some preferential direction (usually, a crystallographic one) and the diffraction spots were blurred. Analysis of many diffraction images revealed that a typical domain size was of the order of 1 μm and the uniaxial part of strain was significantly smaller than its isotropic component and never exceeded ≈2%. The annealing of membranes without free boundaries also led to complete recovery of the original periodicity in both stretched and compressed domains.

2. Binding of Hydrogen to Ripples

In the main text, we have presented a simple argument that atomic hydrogen should preferentially bind to apexes of corrugated graphene due to a contribution from elastic energy. However, there is also a contribution from the electronic energy which works in parallel and promotes the local bonding of atomic hydrogen. The electronic structure of a non-hydrogenated curved graphene is characterized by the appearance of so-called mid-gap states *(S3)* (see the red curve in Figure S2). Our numerical simulations show that the attachment of a pair of hydrogen atoms leads to the splitting of the mid-gap peak in the density of states, leading to the formation of symmetric (donor and acceptor) quasi-localized states (blue curve). For denser hydrogen coverage, an energy gap opens in the electronic spectrum of rippled graphene (green curve). One can see that the splitting moves the mid-gap states away from zero energy, leading to a gain in the total energy. This mechanism seems to be rather general: humps in graphene can stabilize themselves by catching impurities states, and also favours hydrogenation of convex regions.

3. Estimate for the Energy Gap Induced by Hydrogenation of Graphene on Substrate

The notion of preferential hydrogenation of humps on a graphene surface can be combined with the concept of graphene sheets being generally rippled *(S4,S5,S6)*. The combination leads to the following scenario for the disordered derivative obtained by single-sided hydrogenation. This derivative is likely to consist of two phases: convex regions decorated with hydrogen and concave areas that remain non-hydrogenated (see Fig. 3A). In the hydrogenated regions, carbon atoms acquire $sp^3$ hybridisation and an energy gap opens, whereas non-hydrogenated regions retain their $sp^2$ metallic character (see the calculations in Fig. S2). The corresponding electronic spectra are schematically shown in Fig. S3B.

The two-phase picture allows a rough estimate for the energy gap in the hydrogenated convex regions. To this end, we assume that at the neutrality point the metallic concave regions are well separated by the gapped regions so that the system as a whole is insulating. By applying gate voltage, we add charge carriers that fill in the localized states in the gapped regions and the Dirac cones in the metallic ones (see Fig. S3B,C). Eventually, the Fermi level reaches the bottom (top) of the conduction (valence) in the hydrogenated regions, thus bringing the whole system into the metallic regime. Experimentally, this occurs at $n \approx 4 \times 10^{12}$ cm$^{-2}$ (see the main text), which corresponds to a shift of the Fermi energy by ~0.25eV. This yields an energy gap of ~0.5eV for the single-sided hydrogenation. This estimate provides a lower bound for the energy gap as it neglects a contribution from the localised states, quantum confinement effects and a smaller capacitance to the gate for the case of microscopic regions.

FIGURES

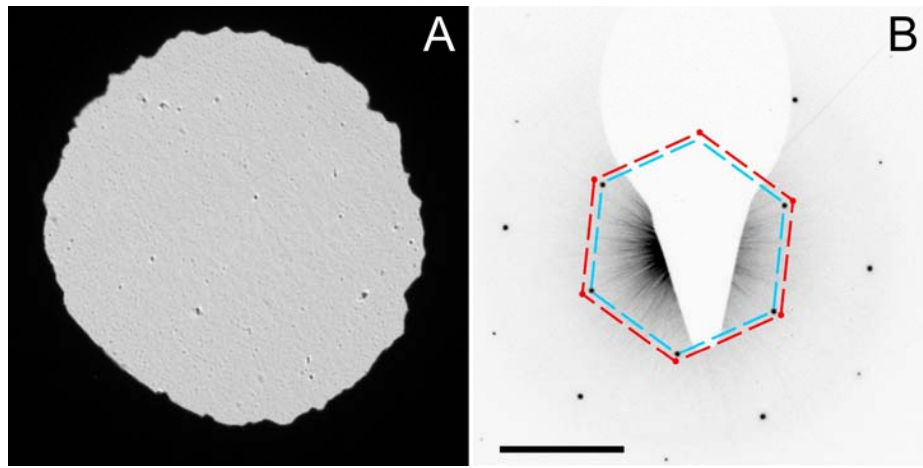

Figure S1. (A) - TEM micrograph of one of our graphene membranes without free boundaries. The presence of a graphene crystal covering the whole aperture is evidenced only by some particulate in the image. (B) - Changes in the lattice constant after extended exposure of this membrane to atomic hydrogen. The scale bar is 5 nm$^{-1}$. The diffraction pattern is for a region with a strongly stretched lattice ($d \approx 2.69$ Å). The diameters of the selected area for the electron diffraction is 0.3 μm. The blue hexagon is a guide to the eye and marks the positions of the diffraction spots. The equivalent spots in unstrained graphene under the same conditions are shown by the red hexagon and dots.

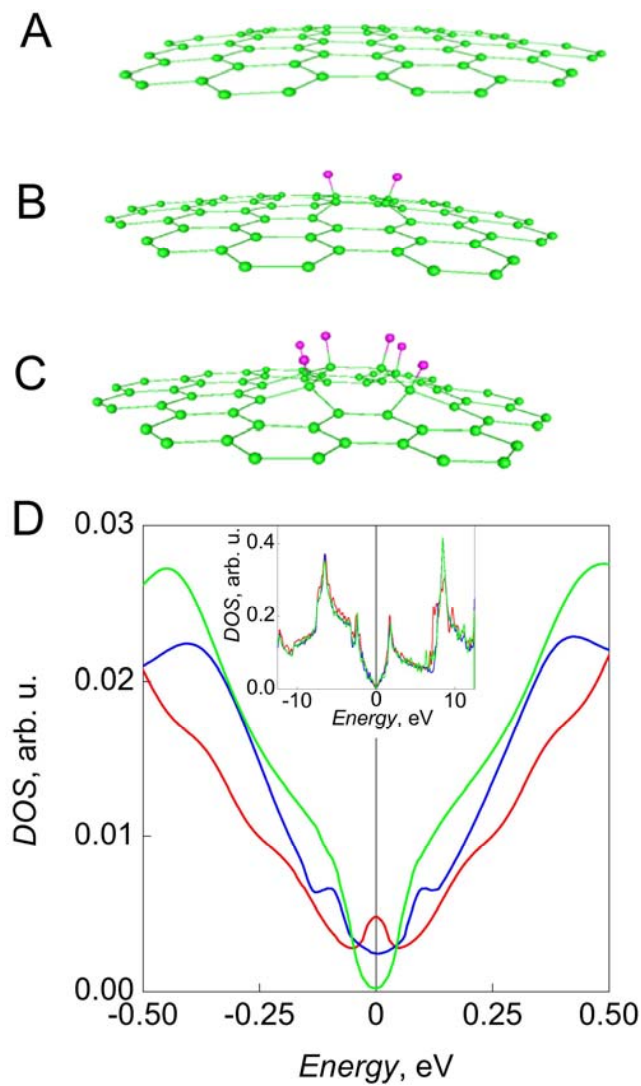

Figure S2. Convex graphene sheet with none (A), two (B) and six (C) hydrogen atoms adsorbed. Atomic coordinates for (B) and (C) were optimized using density functional calculations with the SIESTA code. Configuration (A) was deduced from (B) by removing hydrogen atoms and eliminating the excess displacement for the two carbon atoms that bound hydrogen. (A) corresponds to a ripple of diameter 1.07 nm and height 0.094 nm. (D) – Electronic density of states for configurations A, B and C (red, blue and green curves, respectively). The mid-gap state at zero energy (A; red) becomes split due to adsorption of two hydrogen atoms (B; blue) and a gap opens if more atoms are attached (C; green). Inset: Same calculations for a wider energy range.

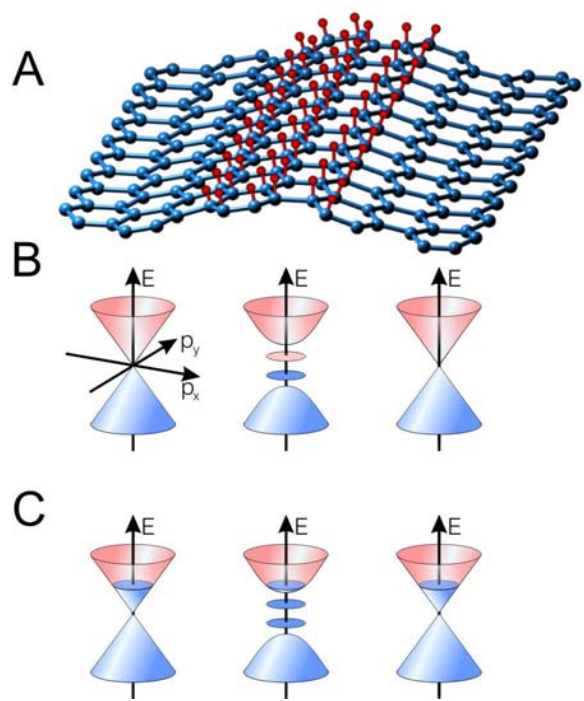

Figure S3. Metal-insulator transition in the disordered graphene derivative obtained by single-sided hydrogenation. (A) – Two-phase model for this derivative: hydrogenated convex regions are adjoined by non-hydrogenated concave ones. Blue (red) spheres represent the carbon (hydrogen) atoms. (B) – Schematic band diagrams for the two phases shown in (A). The diagrams are positioned under the corresponding graphene regions. Hydrogenated regions are represented by a gapped spectrum whereas the concave regions are assumed to be gapless (these are simplified versions of the spectra shown by the red and green curves Fig. S2D). The occupied (unoccupied) states are indicated by blue (pink). The ellipsoids inside the gap represent localised states. The Fermi level in (B) is at the neutrality point. (C) – Same as (B) but the system is doped by electrons so that the Fermi level reaches the bottom of the conductance band in the hydrogenated region.